\documentstyle[amssymb,12pt]{article}

\begin{document}

\title{KdV Surfaces}

\vspace{5cm}

\author{ Metin G{\" u}rses and S{\" u}leyman Tek\\
{\small  Department of Mathematics, Faculty of Sciences}\\
{\small Bilkent University, 06800 Ankara, Turkey}\\}

\vspace{5cm}

\begin{titlepage}
\maketitle

\begin{abstract}
We consider $2$-surfaces arising from the Korteweg de Vries (KdV)
equation. The surfaces corresponding to KdV are in a three
dimensional Minkowski space. They contain a family of quadratic
Weingarten and Willmore-like surfaces. We show that a subset of
KdV surfaces can  be obtained from a variational principle where
the Lagrange function is a polynomial function of the Gaussian and
mean curvatures. We finally  give a method for constructing the
surfaces explicitly, i.e., finding their parametrizations or
finding their position vectors.
\end{abstract}

\end{titlepage}

\section{Introduction}

$2$-surfaces in  ${\mathbb R}^3$ have some special subclasses,
like surfaces of constant Gaussian curvature, surfaces of constant
mean curvature, minimal surfaces, developable surfaces, Bianchi
surfaces, surfaces where the inverse of the mean curvature is
harmonic and the Willmore surfaces. These surfaces arise in many
different branches of sciences; in particular, in various parts of
theoretical physics (string theory, general theory of relativity),
biology and differential geometry \cite{part}, \cite{qu}.

Examples of some of these surfaces like Bianchi surfaces, surfaces
where the inverse of the mean curvature is harmonic \cite{bob},
and the Willmore surfaces \cite{will1}, \cite{will2} are very
rare. The main reason is the difficulty of solving corresponding
differential equations. For this purpose, some indirect methods
\cite{sym1}-\cite{cies2} have been developed for the construction
of $2$- surfaces in ${\mathbb R}^3$ and in three dimensional
Minkowskian geometries ($M_{3}$). Among these methods, soliton
surface technique is very effective. In this method, one mainly
uses the deformations of the Lax equations of the integrable
equations. This way, it is possible to construct families of
surfaces corresponding to some integrable equations like sine
Gordon, Korteweg de Vries (KdV) equation, modified Korteweg de
Vries (mKdV) equation and Nonlinear Schr{\"o}dinger (NLS) equation
\cite{sym1}-\cite{cey},  belonging to the afore mentioned
subclasses of $2$-surfaces in a three dimensional flat geometry.
In particular, using the symmetries of the integrable equations
and their Lax equation, we arrive at classes of $2$-surfaces.
There are many attempts in this direction and examples of new two
surfaces. On the other hand, there are some $2$-surfaces which do
not have many explicit examples and could not be generated by the
solitonic techniques.

Other examples of surfaces are the minimal surfaces \cite{do},
surfaces with constant mean curvature, Willmore surfaces
\cite{will1}, \cite{will2} and surfaces solving the shape equation
\cite{hel1}-\cite{tu3}. All these surfaces come from a variational
principle where the Lagrange function is a polynomial of degree
two in the mean curvature of the surface. There are more general
surfaces solving the Euler-Lagrange equations corresponding to
more general Lagrange functions of the mean and Gaussian
curvatures of the surface \cite{tu1}-\cite{tu3}.

 In this work, by use of the deformation of Lax equations,
we generate some new  Weingarten  and Willmore-like surfaces. For
this purpose, we use the KdV equation and its Lax representation
in $sl(2,R)$ algebra (surfaces in $M_{3}$). In Section 3, we study
the variation of a functional where the Lagrange function is a
function of the mean and Gaussian curvatures. Following
\cite{tu1}-\cite{tu3}, we give the corresponding Euler-Lagrange
equations. Solutions of these equations define a family of
surfaces extremizing the functional we started with. In Section 4,
we give the surfaces corresponding to the KdV equation. These
surfaces contain quadratic Weingarten and Willmore-like surfaces.
We show that KdV surfaces contain also a subclass of surfaces
which extremize families of functionals. For all these surfaces,
we find all possible functionals where the Euler-Lagrange
equations are exactly solved.

Using the method deformation of Lax equations, we can obtain the
fundamental forms, Gauss and mean curvatures of the surfaces. A
parametrization or the position vector of these surfaces can not
be obtained directly. Deformation technique does not produce the
surfaces explicitly. This method reduces the construction of the
surfaces to linear equations (the Lax equations). The solutions of
these linear equations give directly the position vectors of the
corresponding surfaces. In the last section we give such a method.
This method gives the position vectors of the KdV surfaces
explicitly.

\section{Deformation of soliton equations}

Surfaces corresponding to integrable equations are called
integrable surfaces and a connection formula, relating integrable
equations to surfaces,
 was first established by Sym \cite{sym1}, \cite{sym2}.
Here, we shall give a brief introduction (following our previous
work \cite{cey}) of  the recent status of the subject and also
give some new results. Below $M_{2}$ and $M_{3}$ are two and three
dimensional pseudo-Riemannian geometries, respectively.

Let $F:\cal{U} \rightarrow$ $M_{3}$ be an immersion of a domain
 ${\cal{U}} \in M_{2}$  into
$M_{3}$. Let $(x,t) \in \cal{U}$. The surface $F(x,t)$ is uniquely
defined up to rigid motions by the first and the second
fundamental forms. Let $N(x,t)$ be the normal vector field defined
at each point of the surface $F(x,t)$. Then the triple
$\{F_{x},F_{t},N \}$ define a basis of $T_{p}(S)$, where $S$ is
the surface parameterized by $F(x,t)$ and $p$ is a point in $S$,
$p \in S$. The motion of the basis on $S$ is characterized by the
Gauss-Weingarten (GW) equations. The compatibility condition of
these equations are the well-known Gauss-Mainardi-Codazzi (GMC)
equations. The GMC equations are coupled nonlinear partial
differential equations for the coefficients $g_{ij}(x,t)$ and
$h_{ij}(x,t)$ of the first and the second fundamental forms,
respectively. For certain particular surfaces, these equations
reduce to a single or to a system of integrable equations. The
correspondence between the GMC equations and the integrable
equations has been studied extensively, see for example
\cite{cey}.

Let us first give the connection between the integrable equations
with  a surface in $M_{3}:$

\vspace{0.3cm}

 \noindent {\bf Theorem 1} (Fokas-Gelfand
\cite{fok1})\, {\it Let $U(x,t; \lambda), V(x,t; \lambda) ,
A(x,t;\lambda), B(x,t;\lambda)$ take values in an algebra ${\cal
G}$ and let them be differentiable functions of $x,t$ and
$\lambda$ in some neighborhood of $M_{2} \times {\mathbb{R}}$.
Assume that these functions satisfy
\[
U_{t}-V_{x}+[U,V]=0 , \label{cur1}
\]
\noindent and
\[
A_{t}-B_{x}+[A,V]+[U,B]=0 \label{eq1}.
\]
\noindent Define $\Phi(x,t; \lambda)$  in a group ${ G}$ and
suppose that $F(x,t; \lambda)$ takes values in the algebra ${\cal
G}$ by the equations
\[
\Phi_{x}=U\, \Phi~~~,~~~\Phi_{t}=V\, \Phi , \label{cur2}
\]
\noindent and
\[
F_{x}=\Phi^{-1}\,A\, \Phi~~,~~ F_{t}=\Phi^{-1}\,B\, \Phi.
\label{eq0}
\]
\noindent Then for each $\lambda$,  $F(x,t;\lambda)$ defines a
2-dimensional surface in ${\mathbb{R}}^3$,
\[
y_{j}=F_{j}(x,t;\lambda)~~ ,~j=1,2,3~~ ,~~F= \sum^{3}_{k=1}
F_{k}\, e_{k}, \label{eq2}
\]
\noindent where $e_{k}$, ($k=1,2,3$) form a basis of ${\cal G}$.
The first and the second fundamental forms of $S$ are

\begin{eqnarray}
(ds_{I})^2& \equiv &g_{ij}\,dx^{i}\,dx^{j}=
<A,A>\, dx^2+2<A,B>\,dx\,dt+<B,B>\,dt^2 , \nonumber\\
(ds_{II})^2& \equiv &h_{ij}\,dx^{i}\,dx^{j} =<A_{x}+[A,U],C>\,
dx^2 \nonumber
\\&&+2<A_{t}+[A,V],C>\,dx\,dt +<B_{t}+[B,V],C>\,dt^2, \label{eq5}
\end{eqnarray}
\noindent where $i,j=1,2$, $x^1=x$ and $x^2=t$, $
<A,B>=-\frac{1}{2}trace(AB)~,~[A,B]=AB-BA, ~ ||A||=\sqrt{|<A,A>|},
$ and $C=\frac{[A,B]}{||[A,B]||}$. A frame on this surface $S$, is
\[
\Phi^{-1}A \Phi~~,~~ \Phi^{-1}B \Phi~~,~~ \Phi^{-1}C
\Phi.\label{eq9}
\]
The Gauss and the mean curvatures of $S$ are given by
$K=det(g^{-1}\,h)~, ~H=\frac{1}{2}trace(g^{-1}\,h)$.}

\vspace{0.4cm}

From now on, subscripts $x$ and $t$ denote the derivatives of the
objects with respect to $x$ and $t$, respectively. Subscript $nx$
means $n$ times $x$ derivative, where $n$ is positive integer.
Given $U$ and $V$, finding $A$ and $B$ from the equation
$A_{t}-B_{x}+[A,V]+[U,B]=0$ is in general a difficult task.
However, there are some deformations which provide $A$ and $B$
directly. As an example of such  deformations, we shall make use
of the $\lambda $ parameter deformations \cite{sym1},\cite{sym2}:

\vspace{0.3cm}

\[
A={\frac {\partial U}{\partial \lambda}}, ~~B=\frac{\partial
V}{\partial \lambda}, ~~ F=\Phi^{-1}\, \frac{\partial
\Phi}{\partial \lambda}.
\]

\vspace{0.3cm}

For the KdV equation the group $G$ is  $SL(2,R)$ and the algebra
${\cal G}$ is $sl(2,R)$ with the base $2 \times 2$ matrices

\begin{equation}
e_{1}=\left( {\begin{array}{ll}
                           1 & 0 \\
                            0 & -1
                            \end{array}}
                            \right),~~
  e_{2}=\left( {\begin{array}{ll}
                           0 & 1 \\
                            1 & 0
                            \end{array}}
                            \right),~~
e_{3}=\left( {\begin{array}{ll}
                           0 & 1 \\
                            -1 & 0
                            \end{array}}
                            \right).
\end{equation}

\vspace{0.3cm}

\noindent It is clear from the construction that, this technique
does not provide the explicit form of the desired surfaces. To
determine these surfaces, we need the position vector $F$ which is
given as $F=\Phi^{-1}\, \frac{\partial \Phi}{\partial \lambda}$.
This requires that the Lax equations $\Phi_{x}=U \Phi$,\,
\,$\Phi_{t}=V \Phi$ must be solved exactly. This means that for a
given solution of the nonlinear equation (KdV) we have to solve
the corresponding Lax equations. Hence, the deformation technique
is a kind of linearization of the construction of some surfaces.

\section{Surfaces from a variational principle}

Let $H$ and $K$ be the mean and the Gaussian curvatures of a
$2$-surface $S$ (either in $M_{3}$ (three dimensional Minkowski
space) or in ${\mathbb R}^3$) then we have the following
definition.

\vspace{0.3cm}

\noindent {\bf Definition 2 } {\it Let $S$ be a $2$-surface with
its Gaussian  ($K$) and mean ($H$) curvatures.  A functional
${\cal F}$ is defined by

\begin{equation}
{\cal F}=\int_{S}\, {\cal E}(H,K) dA+p \int_{V} dV
\end{equation}

\noindent where ${\cal E}$ is some function of $H$ and $K$, $p$ is
a constant  and $V$ is the volume enclosed within the surface $S$.
For open surfaces, we let $p=0$. }

\vspace{0.3cm}

\noindent The following proposition gives the first variation of
the functional ${\cal F}$.

\vspace{0.3cm}

\noindent {\bf Proposition 3} {\it Let ${\cal E}$ be a twice
differentiable function of $H$ and $K$. Then the Euler-Lagrange
equation for ${\cal F}$ reduces to \cite{tu1}-\cite{tu3}

\begin{equation}\label{el1}
(\nabla^2 +4H^2-2K) {\partial {\cal E} \over \partial
H}+2({\nabla} \cdot \bar{\nabla}+2 K H) {\partial {\cal E} \over
\partial K}-4H {\cal E} +2p=0.
\end{equation}
 Here, and from now on, $\nabla^2={\frac{1}{\sqrt{g}}}{\frac{\partial}{\partial x^i}}
 \left({\sqrt{g}}{g^{ij}}{\frac{\partial}{\partial
 x^j}}\right)$and $\nabla{\cdot}{\bar{\nabla}}={\frac{1}{\sqrt{g}}}{\frac{\partial}{\partial x^i}}
 \left({\sqrt{g}}K{h^{ij}}{\frac{\partial}{\partial
 x^j}}\right)$, ${g}={\det{(g_{ij})}}$, $g^{ij}$ and $h^{ij}$ are
 inverse components of the first and second fundamental forms;
 $x^{i}=(x,t)$ and we assume Einstein's summation convention on
 repeated indices over their ranges.
}

\vspace{0.4cm}

\noindent {\bf Example 1}.\, We have some
examples:\\\begin{itemize}
    \item [{\bf i)}]Minimal surfaces: $ {\cal E}=1, ~~p=0$.
    \item [{\bf ii)}]Constant mean curvature surfaces: ${\cal E}=1$.
    \item [{\bf iii)}]Linear Weingarten surfaces: ${\cal E}=aH+b$, where $a$
and $b$ are some constants.
    \item [{\bf iv)}]Willmore surfaces: $ {\cal E}=H^2$, \cite{will1}, \cite{will2}.
    \item [{\bf v)}]Surfaces solving the shape equation: ${\cal E}=(H-c)^2$,
where $c$ is a constant, \cite{hel1}-\cite{tu3}.
\end{itemize}

\vspace{0.3cm}

\noindent {\bf Definition 4}\, {\it Surfaces obtained from the
solutions of the equation

\begin{equation}
      \nabla^2H+{a}H^3+bH\,K=0,
\end{equation}
where $a$ and $b$ are arbitrary constants, are called {\it
Willmore-like} surfaces.}

\vspace{0.3cm}

\noindent {\bf Remark 1}\, If $a \ne 2$ and $b \ne -2$, then these
surfaces do not arise from a variational problem. The case
$a=-b=2$ corresponds to the Willmore surfaces.

\vspace{0.3cm}

\noindent For compact 2-surfaces, the constant $p$ may be
different than zero, but for noncompact surfaces we assume it to
be zero. For such cases, we require asymptotic conditions, where
$H$ goes to a constant value and $K$ goes to zero asymptotically.
This requires that the KdV  equation must have solutions decaying
rapidly to zero at $|x| \rightarrow \pm \infty$. We know that the
soliton  solutions satisfy this condition. For this purpose, we
shall use the Euler-Lagrange equations (\ref{el1}) for surfaces
obtained by KdV equation and look for solutions (surfaces) of
these equations.

\section{KdV Surfaces}
In this section, we investigate some  surfaces arising from the
KdV equation. KdV surfaces are embedded in a three dimensional
Minkowski space with signature $+1$. We can use our surface
generating technique introduced in Theorem 1.

\vspace{0.5 cm}

\subsection{KdV surfaces from deformations of symmetries}

\vspace{0.3cm} \noindent {\bf Proposition 5}{\label{lamda def1}}
\,{\it Let $u(x,t)$ satisfy

 \begin{equation}{\label{kdv}}
      u_{t}={\frac{1}{4}}u_{3x}+{\frac{3}{2}}uu_{x},
  \end{equation}
and the corresponding ${sl}(2,R)$ valued Lax pair $U,V$ be
 \begin{equation}{\label{U}}
      U=\left(%
            \begin{array}{cc}
                0 & 1 \\
             \lambda-u & 0 \\
            \end{array}%
        \right),
  \end{equation}

 \begin{equation} \label{V}
     V=\left(%
            \begin{array}{cc}
               -\frac{1}{4}u_{x}  & \frac{1}{2}u+\lambda  \\
              -\frac{1}{4}u_{2x}  +\frac{1}{2}\,
               \left( 2\,\lambda +u   \right)
               \left( \lambda-u
              \right) & \frac{1}{4}u_{x} \\
            \end{array}%
       \right),
 \end{equation}
where $\lambda$ is a constant. The corresponding matrices of $U$
and $V$ are
 \begin{equation} \label{A}
     A=\left(%
          \begin{array}{cc}
                 0 & 0 \\
                \mu & 0 \\
          \end{array}%
    \right),
 \end{equation}

 \begin{equation} \label{B}
     B=\left(%
          \begin{array}{cc}
                 0 & \mu \\
                 \frac{\mu}{2}(4\lambda-u) & 0 \\
          \end{array}%
    \right),
 \end{equation}
where $A=\mu(\partial U/\partial\lambda)$, $B=\mu(\partial
V/\partial\lambda)$ and $\mu$ is a constant.

The surface $S$, generated by $U,V,A$ and $B$, has the following
first and second fundamental forms ($i,j=1,2$)

 \begin{eqnarray}
      &&(ds_{I})^{2}=g_{ij}dx^{i}dx^{j}=-\mu^2 dxdt- \frac{\mu^2}{2}(4
       \lambda-u)dt^2,\\
      &&(ds_{II})^{2}=h_{ij}dx^idx^j=-\mu dx^2-{{\mu}}(2
        \lambda+u)dxdt
        -\frac{\mu}{4}\Big(u_{2x}+(u+2\lambda)^2\Big)dt^2,\nonumber
 \end{eqnarray}
with the corresponding Gaussian and mean curvatures
$$K=-\frac{u_{2x}}{\mu^2}, ~~
H=-\frac{2(\lambda-u)}{\mu},$$ where $x^1=x,$ $x^2=t.$}

  \vspace{0.3 cm}

We shall now consider the travelling wave solutions of the KdV
equation. This means that $u_{t}+ u_{x}/c=0$, where $c \ne 0$ is a
constant. Combining this with the KdV equation (\ref{kdv}), we get

 \vspace{0.3 cm}

 \noindent {\bf Proposition 6}
     \,{\it Let $S$ be the surface obtained in Proposition 5
     and $u$ satisfy
  \begin{equation}
      u_{2x}=-{{3}}u^2-{\frac{4}{c}}u+4\beta.
  \end{equation}
Then $S$ is a quadratic Weingarten surface satisfying the relation
 \begin{equation}
      4{\mu^2}cK-\left(\mu H+2\lambda\right)\left[3c\left(\mu
      H+2\lambda\right)+8\right]+16c\beta=0,
 \end{equation}
     where $c \ne 0$ and $\beta$ are constants.}

 \vspace{0.3 cm}

\noindent {\bf Proposition 7}
 \,{\it The surface $S$ defined in Proposition 5
is a Willmore-like surface, i.e. Gaussian and
 mean curvatures satisfy the equation

  \begin{equation}
      \nabla^2H+{a}H^3+bH\,K=0,
  \end{equation}
 where $\nabla^2$ is the Laplace-Beltrami operator on the surface.
 Here, we used the travelling wave solution of KdV equation
 and its consequence $u_{x}^2=-2u^3
 +4\alpha u^2+8\beta u+2\gamma,$
 where
      \begin{equation}
        b=-1,
        a=-\frac{7}{4},
      \end{equation}

      \begin{equation}
        \beta=\frac{28\lambda\alpha-16\alpha^2-21\lambda^2}{20},
     \end{equation}

     \begin{equation}
        \gamma=\frac{16\alpha^3-56\lambda\alpha^2+70\alpha\lambda^2-28\lambda^3}{5}.
     \end{equation}
$\alpha=-{1/c} $,  ($c \ne 0$), $\lambda $ and $c$ are arbitrary
constants.}

\vspace{0.3 cm}

\noindent {\bf Proposition 8}
 \,{\it By using the travelling wave solution of the KdV equation and
  Proposition 5, one can show that the mean curvature of the KdV surface
  $S$ satisfies a more general differential equation

  \begin{eqnarray}
     \nabla^2{H}&=&\frac{1}{2\mu^3}\Big[5\mu^3 H^{3}+2\mu^2(3\lambda-
     2\alpha)H^{2} \nonumber\\
     &&+4\mu(-9\lambda^2+12\alpha \lambda-8\alpha^2
     -12\beta)H\nonumber\\
     &&-56\lambda^3+112\lambda^2 \alpha
     -64\alpha^2\lambda+32\lambda \beta
     -64\alpha \beta-16\gamma\Big].
  \end{eqnarray}}

 \vspace{0.3 cm}

\subsection{KdV surfaces from a variational principle}

\vspace{0.3 cm}

 \noindent {\bf Proposition 9}
 \,{\it KdV surface $S$
defined in Proposition 5 satisfies
 the generalized shape equation (\ref{el1})
for some ${\cal E}$, which is a polynomial of $H$ and $K$. Let
deg$({\cal E})=N$, then
\begin{itemize}
    \item [i)] for $N=3:$
     \subitem ${\cal
     E}=a_{1}H^3+a_{2}H^2+a_{3}H+a_{4}+a_{5}K+a_{6}KH$,
    \subsubitem$a_{1}=
     \displaystyle-{8165p\mu^4}\Big/
     3\Big(54944\alpha\gamma-553728\lambda\alpha^3-2558912\lambda^3\alpha
     +1420444\lambda\alpha\beta+3338368\lambda^2\alpha^2-416080\lambda^2\beta
     -19456\lambda\gamma+407552\lambda^4+95427\beta^2+94848\alpha^2\beta\Big)$,
    \vspace{0.1 cm}
    \subsubitem$a_{2}=
    \displaystyle\frac{6a_{1}(14533\lambda-12312\alpha)}{8165\mu}$,
    \vspace{0.1 cm}
    \subsubitem$a_{3}=-
    \displaystyle\frac{9a_{1}(72064\alpha^2+187636\lambda^2-331168\lambda\alpha
    -108813\beta)}{8165\mu^2}$,
    \vspace{0.1 cm}
    \subsubitem $a_{4}=
    \displaystyle a_{1}\Big(523816\lambda^3-2214672\lambda^2\alpha-415296\alpha^3
    +2503776\lambda\alpha^2+1018557\alpha\beta-304368\lambda\beta+27600\gamma\Big)\Big/
    {8165\mu^3},$
    \vspace{0.1 cm}
    \subsubitem $a_{6}=
    \displaystyle-\frac{972a_{1}}{355},$

    \vspace{0.1 cm}

    where $\mu\neq0,$ $p\neq0,$ $\lambda,$ $\alpha =-{1/c}$, ($c \ne 0$), $\beta,$ $\gamma$ and $a_{5}$
    are arbitrary constants, but $\lambda,$ $\alpha,$ $\beta$ and
    $\gamma$ can not be zero at the same time.
    \vspace{0.3cm}
    \item [ii)] for $N=4:$
    \subitem ${\cal E}=a_{1}H^4+a_{2}H^3+a_{3}H^2+a_{4}H+a_{5}
    +a_{6}K+a_{7}KH+a_{8}KH^2,$
    \subsubitem $a_{{2}}=\Big(\bigg[
    -303319296\,{\lambda}^{2}\gamma-502986720\,\gamma\,\beta-
10818223680\,{\lambda}^{3}\beta-4217392512\,\beta\,{\alpha}^{3}-
22696872768\,\lambda\,{\alpha}^{2}\beta+3385990152\,\lambda\,{\beta}^{
2}-4991385168\,\alpha\,{\beta}^{2}-1902053376\,{\alpha}^{2}\gamma-
12134983680\,{\lambda}^{2}{\alpha}^{3}+36770221056\,{\lambda}^{3}{
\alpha}^{2}-6024089088\,\lambda\,{\alpha}^{4}-24364181760\,{\lambda}^{
4}\alpha+31860385824\,{\lambda}^{2}\alpha\,\beta+1157755008\,\lambda\,
\alpha\,\gamma+5069893632\,{\lambda}^{5} \bigg] a_{{1}}-9250945
\,p{\mu}^{5}\Big)\Big/\Big(3399\,\mu\, \bigg[
-553728\,\lambda\,{\alpha}^{3}+3338368\,{\lambda}^{
2}{\alpha}^{2}-416080\,{\lambda}^{2}\beta+95427\,{\beta}^{2}+94848\,{
\alpha}^{2}\beta-2558912\,{\lambda}^{3}\alpha-19456\,\lambda\,\gamma+
407552\,{\lambda}^{4}+1420444\,\lambda\,\alpha\,\beta+54944\,\alpha\,
\gamma \bigg] \Big)$,
     \vspace{0.1 cm}
    \subsubitem $a_{{3}}=\displaystyle
   - 6\Big(\bigg[-14533\,\lambda+12312\,\alpha  \bigg]1133\,\mu\, a_{{2}}+
 \bigg[63409476\,{\lambda}^{2}+66459936\,{\alpha}^{2}+40215820\,\beta
-107133888\,\lambda\,\alpha \bigg]
a_{{1}}\Big)\Big/\Big(9250945\mu^2\Big)$,
    \vspace{0.1 cm}
    \subsubitem $a_{{4}}=\displaystyle
   \Big(-\bigg[-108813\,\beta+187636\,{\lambda}^{2}-331168\,
\lambda\,\alpha+72064\,{\alpha}^{2} \bigg] 10197\,\mu\,a_{{2}}+
\bigg[ 2257768448\,{\lambda}^{3}+359782560\,\gamma-1687519464
\,\lambda\,\beta+602232192\,{\alpha}^{3}-6679109376\,{\lambda}^{2}
\alpha+4236473088\,\lambda\,{\alpha}^{2}+4082502096\,\alpha\,\beta
 \bigg] a_{{1}}\Big)\Big/\Big(9250945\mu^3\Big)$,
    \vspace{0.1 cm}
    \subsubitem $a_{{5}}=\displaystyle-
   \Big(\Big[-523816\,{\lambda}^{3}+415296\,{\alpha}^{3}+304368
\,\lambda\,\beta-27600\,\gamma-1018557\,\alpha\,\beta-2503776\,\lambda
\,{\alpha}^{2}+2214672\,{\lambda}^{2}\alpha \Big]1133\,\mu\,
a_{2}+\Big[-757178400\,\alpha\,\gamma+2345255472\,{\lambda}^{4}+8043002568\,
\lambda\,\alpha\,\beta-1506022272\,{\alpha}^{4}-3033745920\,\lambda\,{
\alpha}^{3}-7014105936\,{\alpha}^{2}\beta-1846341216\,{\lambda}^{2}
\beta+7872480\,{\beta}^{2}+543057600\,\lambda\,\gamma+10787593728\,{
\lambda}^{2}{\alpha}^{2}-8662258752\,{\lambda}^{3}\alpha\Big]a_{1}\Big)\Big/
\Big(9250945\mu^4\Big)$,
     \vspace{0.1 cm}
    \subsubitem $a_{{7}}=\displaystyle
    {\frac {12\Big([ -517728\,\alpha+670472\,\lambda ]a_{{1}}-91773\,a_{{2}}
\mu\Big)}{402215\mu}}$,
    \vspace{0.1 cm}
    \subsubitem $a_{8}=\displaystyle-\frac{2280a_{1}}{1133}$,

     \vspace{0.1 cm}

  where $\mu\neq0,$ $\lambda,$ $\alpha=-{1/c}$, ($c \ne 0$), $\beta,$
    $\gamma$, $a_{1}$, $a_{6}$ and $p$
     are arbitrary constants, but $\lambda,$ $\alpha,$ $\beta$ and
    $\gamma$ can not be zero at the same time.
     \vspace{0.3cm}
\end{itemize}
 Here we used the travelling wave solution
 $u_{t}=\alpha u_{x},~(\alpha=-\displaystyle 1/c)$ of the KdV equation
 and its consequences
 $u_{x}^2=-2u^3+4\alpha u^2+8\beta u+2\gamma,$
 $u_{2x}=-{3}u^2+4\alpha u+4\beta,$
 $u_{4x}=-6u_{x}^2+(4\alpha-6u)u_{2x}$.}

 \vspace{0.3 cm}

\noindent {\bf Remark 2}
 \,{\it   The KdV surface satisfies the Euler-Lagrange
equation
      (\ref{el1}) for the Lagrangian with degree $N$
  \begin{equation}
       {\cal E}=\sum_{n=1}^{N}\Big(\sum_{k=1}^{N-n}
       a_{kn}H^{2k+1}\Big)K^{n}+
       \sum_{l=0}^{N}b_{l}H^{l}+eK,~~N=3,4,5,...
  \end{equation}
 Here $a_{kn},$ $b_{l}$ and $e$ are constants. Some of the
 constants can be written in terms of others.}

\vspace {0.3 cm}

\section{Derivation of the surfaces}

\vspace{0.3cm}

 In the previous sections, we found possible surfaces satisfying
 certain equations, without giving the  $F$ functions explicitly. In
 this section, we shall find the position vector
 $\overrightarrow{y}=\left(y_{1}(x,t),y_{2}(x,t),y_{3}(x,t)\right)$
 of the corresponding KdV surfaces. To determine $\overrightarrow{y}$, we use
 the equations
  \begin{equation}
  F_{x}={\Phi^{-1}}A\Phi,~~ F_{t}={\Phi^{-1}}B\Phi,
  \end{equation}
where $F=\overrightarrow{\sigma}{\cdot}\overrightarrow{y}.$ Hence,
we need the $2\times 2$ matrix $\Phi$ solving the Lax equation for
the given function $u(x,t).$ Our method of constructing the
position vector $\overrightarrow{y}$ of integrable surfaces
consists of the following steps:

 \vspace{0.3 cm}

 {\noindent}{\bf{i)}} Finding a solution $u=u(x,t)$ of the KdV equation
 with a given symmetry:

 \vspace{0.1 cm}
\noindent Here, we consider travelling wave solutions
 $u_{t}=-u_{x}/c$. By using this assumption we get
  \begin{equation}
  u_{x}^2=-2u^3-{\frac{4}{c}}u^2+8\beta u+2\gamma,
  \end{equation}
where $c\neq 0,$ $\beta$ and $\gamma$ are arbitrary constants.

\vspace{0.3 cm}

{\noindent}{\bf{ii)}} Finding solution of the Lax equation
 \begin{equation}
 \Phi_{x}=U\Phi,~~\Phi_{t}=V\Phi,
 \end{equation}
for given $U$ and $V$:

\vspace{0.1 cm} \noindent In our case, corresponding ${sl}(2,R)$
valued $U,$ $V$ of the KdV equation are given in (\ref{U}) and
(\ref{V}).
 Consider the $2\times 2$ matrix $\Phi$
 \begin{equation}
 \Phi=\left(%
\begin{array}{cc}
  \Phi_{11} & \Phi_{12} \\
  \Phi_{21} & \Phi_{22} \\
\end{array}%
\right).
 \end{equation}
By using $\Phi$ and $U$, we can write $\Phi_{x}=U\Phi$ in matrix
form as
 \begin{equation}
 \left(%
\begin{array}{cc}
  (\Phi_{11})_{x} & (\Phi_{12})_{x} \\
  (\Phi_{21})_{x} & (\Phi_{22})_{x} \\
\end{array}%
\right)=\left(%
\begin{array}{cc}
  \Phi_{21} & \Phi_{22} \\
  (\lambda-u)\Phi_{11} & (\lambda-u)\Phi_{12} \\
\end{array}%
\right).
 \end{equation}
By using $(\Phi_{11})_{x}=\Phi_{21}$ and
$(\Phi_{21})_{x}=(\lambda-u)\Phi_{11}$, we have
 \begin{equation}{\label{psi_11xx}}
 (\Phi_{11})_{xx}-(\lambda-u)\Phi_{11}=0.
 \end{equation}
Similarly, we have an equation for $\Phi_{12}$ as
 \begin{equation}{\label{psi_12xx}}
 (\Phi_{12})_{xx}-(\lambda-u)\Phi_{12}=0.
 \end{equation}
By solving (\ref{psi_11xx}) and (\ref{psi_12xx}) we determine the
explicit $x$-dependence of $\Phi_{11}, \Phi_{12}$ and also
$\Phi_{21}, \Phi_{22}$. By using $\Phi_{t}=V\Phi$, we get
 \begin{eqnarray}
&&\label{psi_11t}(\Phi_{11})_{t}=-\frac{1}{4}u_{x}\Phi_{11}+(\frac{1}{2}u+\lambda)\Phi_{21},\\
&&\label{psi_21t}(\Phi_{21})_{t}=\left[-\frac{1}{4}u_{2x}
+\frac{1}{2}\,
               \left( 2\,\lambda +u   \right)
               \left( \lambda-u
              \right)\right]\Phi_{11}+\frac{1}{4}u_{x}\Phi_{21},
 \end{eqnarray}
and
 \begin{eqnarray}
&&\label{psi_12t}(\Phi_{12})_{t}=-\frac{1}{4}u_{x}\Phi_{12}+(\frac{1}{2}u+\lambda)\Phi_{22},\\
&&\label{psi_22t}(\Phi_{22})_{t}=\left[-\frac{1}{4}u_{2x}
+\frac{1}{2}\,
               \left( 2\,\lambda +u   \right)
               \left( \lambda-u
              \right)\right]\Phi_{12}+\frac{1}{4}u_{x}\Phi_{22}.
 \end{eqnarray}
By solving these equations, we determine the explicit
$t$-dependence of $\Phi_{11},$ $\Phi_{21},$ $\Phi_{12}$ and
$\Phi_{22.}$ Hence we complete finding the solution $\Phi$ of the
Lax equation.

\vspace{0.3 cm}

{\noindent}{\bf{iii)}} Finding $F$:

\vspace{0.1 cm} \noindent If $\Phi$ depends on $\lambda$
explicitly, $F$ can be found directly from
 \begin{equation}
 F={\Phi^{-1}}{\frac{\partial \Phi}{\partial
 \lambda}}=y_{1}e_{1}+y_{2}e_{2}+y_{3}e_{3}.
 \end{equation}
If $\Phi$ has been found for a fixed value of $\lambda$, we use
 $F_{x}={\Phi^{-1}}A\Phi,~~ F_{t}={\Phi^{-1}}B\Phi$ to find
 $F=y_{1}e_{1}+y_{2}e_{2}+y_{3}e_{3}$. For our case, $A$ and $B$
 are given in (\ref{A}) and (\ref{B}) which are
 the corresponding matrices of $U$ and $V$. Integrating the equations

\begin{equation}
F_{x}={\Phi^{-1}}A\Phi,~~ F_{t}={\Phi^{-1}}B\Phi,
\end{equation}
we get $F$. By writing $F$ as a linear combination of $e_{1},$
$e_{2}$ and $e_{3}$, and collecting the coefficients of $e_{i}$
($i=1,2,3$), we get the components of the vector
$\overrightarrow{y}.$

\vspace{0.3 cm}

\noindent {\bf Example 2}.\, Let $u=u_{0}=\frac{2}{3}(\alpha \pm
\sqrt{\alpha^2+3\beta})$ be the constant solution of the
integrated form $u_{x}^2+2u^3-4\alpha u^2-8\beta
 u-2\gamma=0$ of the KdV equation
 $u_{t}={\frac{1}{4}}u_{3x}+{\frac{2}{3}}uu_{x},$ where $\alpha=-{1}/{c},$ $c \neq
 0$. By denoting $\lambda-u_{0}=m^2$, we find
 the solutions of (\ref{psi_11xx}) and (\ref{psi_12xx}) as
  \begin{equation}{\label{ppsi11(x)}}
  \Phi_{11}=A_{1}(t)e^{{mx}}+B_{1}(t)e^{-mx},
  \end{equation}
 \begin{equation}{\label{ppsi12(x)}}
   \Phi_{12}=A_{2}(t)e^{{mx}}+B_{2}(t)e^{-mx},
  \end{equation}
and
 \begin{equation}{\label{ppsi21(x)}}
 \Phi_{21}=(\Phi_{11})_{x}=m\,[A_{1}(t)e^{{mx}}-B_{1}(t)e^{-mx}],
  \end{equation}
 \begin{equation}{\label{ppsi22(x)}}
  \Phi_{22}=(\Phi_{12})_{x}=m\,[A_{2}(t)e^{{mx}}-B_{2}(t)e^{-mx}].
  \end{equation}
Since $u$ is constant,
\begin{equation}{\label{Phi_11t}}
(\Phi_{11})_{t}=(\frac{1}{2}u_{0}+\lambda)\Phi_{21},~~
(\Phi_{21})_{t}=\left[\frac{1}{2}\,
               \left( 2\,\lambda +u_{0} \right)m\right]\Phi_{11},
 \end{equation}
and
 \begin{equation}{\label{Phi_12t}}
(\Phi_{12})_{t}=(\frac{1}{2}u_{0}+\lambda)\Phi_{22},~~
(\Phi_{22})_{t}=\left[\frac{1}{2}\,
               \left( 2\,\lambda +u_{0}\right)m\right]\Phi_{12}.
 \end{equation}
Denoting $\frac{1}{2}(2\lambda+u_{0})=n$ and using
(\ref{ppsi11(x)}), (\ref{ppsi12(x)}), (\ref{ppsi21(x)}) and
(\ref{ppsi22(x)}) in the equations (\ref{Phi_11t}) and
(\ref{Phi_12t}), we find
 \begin{equation}
 A_{1}(t)=C_{1}e^{n{m}t},~~B_{1}(t)=D_{1}e^{-n{m}t},
 \end{equation}
\begin{equation}
 A_{2}(t)=C_{2}e^{n{m}t},~~B_{2}(t)=D_{2}e^{-n{m}t},
 \end{equation}
where $C_{1},$ $C_{2}$, $D_{1}$ and $D_{2}$ are arbitrary
constants. Thus $\Phi$ has the following form
 \begin{equation}{\label{Phi for cons sol}}
 \Phi=\left(%
\begin{array}{cc}
  C_{1}e^{m(nt+x)}+D_{1}e^{-m(nt+x)} &
  C_{2}e^{m(nt+x)}+D_{2}e^{-m(nt+x)} \\
  {m}( C_{1}e^{m(nt+x)}-D_{1}e^{-m(nt+x)}) &
  {m}( C_{2}e^{m(nt+x)}-D_{2}e^{-m(nt+x)}) \\
\end{array}%
\right).
 \end{equation}

 \vspace{0.3 cm}

\noindent{\bf Remark 3}
 \,{\it$\det(\Phi)=2m(C_{2}D_{1}-C_{1}D_{2})=$ constant, in Example 2.}

 \vspace{0.3 cm}

 {\noindent}Since $\Phi$ depends on $\lambda$ explicitly in (\ref{Phi for cons sol}), we can write $F$ directly as
 \begin{equation}
 F={\Phi^{-1}}{\frac{\partial \Phi}{\partial
 \lambda}}=y_{1}e_{1}+
 y_{2}e_{2}+
 y_{3}e_{3},
 \end{equation}
where $e_{1},e_{2},e_{3}$ are basis elements of $sl(2,R)$ and
 \begin{eqnarray}
 &&y_{1}=-\left(\frac{D_{1}C_{2}+C_{1}D_{2}}{D_{1}C_{2}-C_{1}D_{2}}\right)
 \frac{(4\lambda-u)t+x}{2\sqrt{\lambda-u}},\label{y1}\\
 &&y_{2}=\left(\frac{D_{1}C_{1}-D_{2}C_{2}}{D_{1}C_{2}-D_{2}C_{1}}\right)
 \frac{(4\lambda-u)t+x}{2\sqrt{\lambda-u}},\label{y2}\\
&&y_{3}=-\left(\frac{D_{1}C_{1}+D_{2}C_{2}}{D_{1}C_{2}-D_{2}C_{1}}\right)
 \frac{(4\lambda-u)t+x}{2\sqrt{\lambda-u}}.\label{y3}
 \end{eqnarray}
Thus we find the position vector
$\overrightarrow{y}=(y_{1}(x,t),y_{2}(x,t),y_{3}(x,t))$, where
$y_{1},$ $y_{2}$ and $y_{3}$ are functions given as (\ref{y1}),
(\ref{y2}) and (\ref{y3}), respectively. This solution corresponds
to a plane in ${\mathbb{R}}^{3}.$

\vspace{0.3 cm}

\noindent {\bf Example 3}.\, Let
$u={2k^2c^2}{{\textrm{sech}}^2k(t-cx)}$ be a solution of the KdV
equation, where $k^2=-{1}/{c^3}$.
 By denoting $k(t-cx)=\xi$, we find
 the solutions of (\ref{psi_11xx}) and (\ref{psi_12xx}) as
  \begin{equation}{\label{psi11(x)}}
  \Phi_{11}=A_{1}(t){{\textrm{sech}}\,\xi}+B_{1}(t)[\sinh
 \xi+\xi\,{{\textrm{sech}}\,\xi}],
  \end{equation}
 \begin{equation}{\label{psi12(x)}}
 \Phi_{12}=A_{2}(t){{\textrm{sech}}\,\xi}+B_{2}(t)[\sinh
 \xi+\xi\,{{\textrm{sech}}\,\xi}],
  \end{equation}
and
 \begin{equation}{\label{psi21(x)}}
 \Phi_{21}=(\Phi_{11})_{x}=kc\,A_{1}(t)\,{{\textrm{sech}}\,\xi}\,{\tanh
\xi}+kc\,B_{1}(t)\,[{\xi}\,{{{\textrm{sech}}\,\xi}}\,{\tanh{\xi}}-{\cosh{\xi}}-{{\textrm{sech}}\,\xi}],
  \end{equation}
 \begin{equation}{\label{psi22(x)}}
  \Phi_{22}=(\Phi_{12})_{x}=kc\,A_{2}(t)\,{{\textrm{sech}}\, \xi}\,{\tanh
\xi}+kc\,B_{2}(t)\,[{\xi}\,{{{\textrm{sech}}\,\xi}}\,{\tanh{\xi}}-{\cosh{\xi}}-{{\textrm{sech}}\,\xi}],
  \end{equation}
for $\lambda=k^2c^2.$ By using these functions and considering
(\ref{psi_11t}), (\ref{psi_12t}), (\ref{psi_21t}), (\ref{psi_22t})
with $u_{x}=4k^3c^3\,{{\textrm{sech}}^2\xi}\,{\tanh{\xi}},$
$u_{2x}=4k^3c^3\left(2{{\textrm{sech}}^2\xi}\,{\tanh^2{\xi}}-{{\textrm{sech}}^4\xi}\right)$,
we get
\begin{equation}
B_{1}(t)=B_{1}~~ {\textrm{and}} ~~A_{1}(t)=2B_{1}kt+C_{1},
 \end{equation}
\begin{equation}
B_{2}(t)=B_{2}~~ {\textrm{and}} ~~A_{2}(t)=2B_{2}kt+C_{2},
 \end{equation}
where $B_{1},$ $B_{2}$, $C_{1}$ and $C_{2}$ are arbitrary
constants. Thus components of $\Phi$ are
 \begin{equation}
 \Phi_{11}=B_{1}\left(2kt\,{\textrm{sech}{\,\xi}}+\sinh{\xi}+
 {\xi}\,{\textrm{sech}{\,\xi}}\right)
 +C_{1}\,{\textrm{sech}{\, \xi}},\label{Phi11}
 \end{equation}
\begin{equation}
\Phi_{12}=B_{2}\left(2kt\,{\textrm{sech}{\,\xi}}+\sinh{\xi}+
{\xi}\,{\textrm{sech}{\,\xi}}\right)
 +C_{2}\,{\textrm{sech}{\, \xi}},\label{Phi12}
\end{equation}
\begin{eqnarray}
 \Phi_{21}&=&kc\bigg[B_{1}\bigg(2kt\,{\textrm{sech}{\,\xi}}\tanh{\xi}
 -\cosh{\xi}-{\textrm{sech}{\,\xi}}+{\xi}\,{\textrm{sech}{\,\xi}}\tanh{\xi}\bigg)\\ \nonumber
&&+C_{1}\,{\textrm{sech}{\,\xi}}\tanh{\xi}\bigg],\label{Phi21}
\end{eqnarray}
\begin{eqnarray}
 \Phi_{22}&=&kc\bigg[B_{2}\bigg(2kt\,{\textrm{sech}{\,\xi}}\tanh{\xi}
 -\cosh{\xi}-{\textrm{sech}{\,\xi}}+{\xi}\,{\textrm{sech}{\,\xi}}\tanh{\xi}\bigg)\\ \nonumber
 &&+C_{2}\,{\textrm{sech}{\,\xi}}\tanh{\xi}\bigg].\label{Phi22}
\end{eqnarray}

\vspace{0.3 cm}

\noindent {\bf Remark 4}
 \,{\it$\det(\Phi)=2kc(C_{2}B_{1}-C_{1}B_{2})=$ constant, in Example 3.}

 \vspace{0.3 cm}

 {\noindent}Since  $\Phi$ is determined for fixed value of $\lambda$, where components are given
 as (\ref{Phi11}), (\ref{Phi12}), (\ref{Phi21}) and (\ref{Phi22}), we obtain $F_{x}$ and $F_{t}$.
 They read
  \begin{equation}
  F_{x}={\Phi^{-1}}A\Phi=\left(%
\begin{array}{cc}
  F_{x}^{11} & F_{x}^{12} \\
 F_{x}^{21} & F_{x}^{22} \\
\end{array}%
\right),~~F_{t}={\Phi^{-1}}B\Phi=\left(%
\begin{array}{cc}
  F_{t}^{11} & F_{t}^{12} \\
  F_{t}^{21} & F_{t}^{22}\\
\end{array}%
\right),
  \end{equation}
where
\begin{eqnarray}
F_{x}^{11}&=&-\frac{\mu}{2kc\cosh^2{\xi}\, \left( C_{{2}}B_{{ 1}}
-C_{{1}}B_{{2}}\right)}\Big[B_{1}B_{2}\Big(2\xi
\sinh{\xi}\cosh{\xi}\\
\nonumber
&&+4kt\sinh{\xi}\cosh{\xi}+\cosh^4{\xi}+4kt{\xi}+{\xi}^2+4k^2t^2-\cosh^2{\xi}\Big)\\
\nonumber
&&+\Big(B_{1}C_{2}+B_{2}C_{1}\Big)\Big(\sinh{\xi}\cosh{\xi}+2kt+\xi\Big)+C_{1}C_{2}\Big],
\end{eqnarray}
\begin{eqnarray}
F_{x}^{12}&=&-\frac{\mu}{2kc\cosh^2{\xi}\,\left( C_{{2}}B_{{ 1}}
-C_{{1}}B_{{2}}\right)}\Big[B_{2}^2\Big(2\xi
\sinh{\xi}\cosh{\xi}\\
\nonumber
&&+4kt\sinh{\xi}\cosh{\xi}+\cosh^4{\xi}+4kt{\xi}+{\xi}^2+4k^2t^2-\cosh^2{\xi}\Big)\\
\nonumber
&&+2B_{2}C_{2}\Big(\sinh{\xi}\cosh{\xi}+2kt+\xi\Big)+C_{2}^2\Big],
\end{eqnarray}
\begin{eqnarray}
F_{x}^{21}&=&-\frac{\mu}{2kc\cosh^2{\xi}\,\left( C_{{2}}B_{{ 1}}
-C_{{1}}B_{{2}}\right)}\Big[B_{1}^2\Big(2\xi
\sinh{\xi}\cosh{\xi}\\
\nonumber
&&+4kt\sinh{\xi}\cosh{\xi}+\cosh^4{\xi}+4kt{\xi}+{\xi}^2+4k^2t^2-\cosh^2{\xi}\Big)\\
\nonumber
&&+2B_{1}C_{1}\Big(\sinh{\xi}\cosh{\xi}+2kt+\xi\Big)+C_{1}^2\Big],
\end{eqnarray}
\begin{equation}
F_{x}^{22}=-F_{x}^{11},
\end{equation}

\begin{eqnarray}
F_{t}^{11}&=&-\frac{\mu}{2kc\cosh^2{\xi}\,\left( C_{{2}}B_{{ 1}}
-C_{{1}}B_{{2}}\right)}\Big[B_{1}B_{2}\Big(6\xi
\sinh{\xi}\cosh{\xi}\\
\nonumber
&&+12kt\sinh{\xi}\cosh{\xi}+\cosh^4{\xi}+4kt{\xi}+4k^2t^2+{\xi}^2-5\cosh^2{\xi}\Big)\\
\nonumber
&&+\Big(B_{1}C_{2}+B_{2}C_{1}\Big)\Big(3\sinh{\xi}\cosh{\xi}+2kt+\xi\Big)+C_{1}C_{2}\Big],
\end{eqnarray}
\begin{eqnarray}
F_{t}^{12}&=&-\frac{\mu}{2kc\cosh^2{\xi}\,\left( C_{{2}}B_{{ 1}}
-C_{{1}}B_{{2}}\right)}\Big[B_{2}^2\Big(6\xi
\sinh{\xi}\cosh{\xi}\\
\nonumber
&&+12kt\sinh{\xi}\cosh{\xi}+\cosh^4{\xi}+4kt{\xi}+4k^2t^2+{\xi}^2-5\cosh^2{\xi}\Big)\\
\nonumber
&&+2B_{2}C_{2}\Big(3\sinh{\xi}\cosh{\xi}+2kt+\xi\Big)+C_{2}^2\Big],
\end{eqnarray}
\begin{eqnarray}
F_{t}^{21}&=&-\frac{\mu}{2kc\cosh^2{\xi}\,\left( C_{{2}}B_{{ 1}}
-C_{{1}}B_{{2}}\right)}\Big[B_{1}^2\Big(6\xi
\sinh{\xi}\cosh{\xi}\\
\nonumber
&&+12kt\sinh{\xi}\cosh{\xi}+\cosh^4{\xi}+4kt{\xi}+4k^2t^2+{\xi}^2-5\cosh^2{\xi}\Big)\\
\nonumber
&&+2B_{1}C_{1}\Big(3\sinh{\xi}\cosh{\xi}+2kt+\xi\Big)+C_{1}^2\Big],
\end{eqnarray}
\begin{equation}
F_{t}^{22}=-F_{t}^{11}.
\end{equation}
By solving these equations, we get the position vector of the
surface through the function $F$ corresponding to the KdV equation
with non-constant solution as
 \begin{equation}
 F=y_{1}e_{1}+y_{2}e_{2}+y_{3}e_{3},
 \end{equation}
where $e_{1},e_{2},e_{3}$ are basis elements of $sl(2,R)$ and
\begin{eqnarray}
&&y_{1}=2\zeta_{1}\left(B_{1}B_{2}\zeta_{2}+\zeta_{3}(C_{1}B_{2}+C_{2}B_{1})+16c^3C_{1}C_{2}\right),\label{y11}\\
&&y_{2}=\zeta_{1}\left(\zeta_{2}(B_{2}^2-B_{1}^2)+2\zeta_{3}(B_{1}C_{1}-B_{2}C_{2})-16c^3(C_{1}^2-C_{2}^2)\right),
\label{y22}\\
&&y_{3}=\zeta_{1}\left(\zeta_{2}(B_{1}^2+B_{2}^2)+2\zeta_{3}(B_{1}C_{1}+B_{2}C_{2})+16c^3(C_{1}^2+C_{2}^2)\right),
\label{y33}\\
 &&\zeta_{1}=\frac{\mu}{32c^2(B_{1}C_{2}-B_{2}C_{1})(1+e^{2\xi})},\\
 &&\zeta_{2}=-8\big(1-e^{2\xi}\big)\big(3t-cx\big)^2+4kc^3(9t-cx)(1+e^{2\xi})\\ \nonumber
 &&~~~~~~+c^3(1-e^{4\xi})-2c^3\sinh{2\xi},\\
&&\zeta_{3}=8kc^3\left(3t-cx\right)\left(1-e^{2\xi}\right).
\end{eqnarray}
 Thus we find the $\overrightarrow{y}=(y_{1}(x,t),y_{2}(x,t),y_{3}(x,t))$ vector,
where $y_{1},$ $y_{2}$ and $y_{3}$ are given as (\ref{y11}),
(\ref{y22}) and (\ref{y33}), respectively.

 \vspace{0.3 cm}

\section{Conclusion}

In this work, we considered two families of surfaces, the
Willmore-like surfaces and the surfaces derivable from a
variational principle. Willmore-like surfaces, except for some
particular values of the parameters, do not arise from a
variational problem. To construct these two families of surfaces,
we introduced a two step procedure. The first step is to use the
method of deformation of the Lax equations corresponding to
nonlinear partial differential equations. Any surface obtained
through this method is called {\it integrable}. At this step, it
is possible to find the first and second fundamental forms, the
Gaussian and mean curvatures of these surfaces. In the second step
of our approach, we determine the explicit locations, i.e., the
position vectors, of these surfaces by solving the corresponding
Lax equations of some integrable equations. As an application we
used the KdV equation and its Lax equation. Corresponding to these
equations, we have found several families of Willmore-like
surfaces and a hierarchy of surfaces arising from a variational
problem, where the Lagrange function is a polynomial of the
Gaussian and mean curvatures of these surfaces.

 \vspace{2cm}
We thank {\" O}zg{\" u}r Sar{\i}o{\~ g}lu for his comments and for
his critical reading of the manuscript.
 This work is partially supported by the Scientific and Technical
 Research Council of Turkey and Turkish Academy of Sciences.

\end{document}